%% file: main.tex
\documentclass{article}
\usepackage{spconf,amsmath,graphicx}
\usepackage[table,x11names,dvipsnames]{xcolor}
\usepackage{etoolbox}
\usepackage{booktabs}
\usepackage{xurl}
\usepackage{tabularx}

\newcommand\clearrow{\global\let\rowmac\relax}
\usepackage{siunitx}
\usepackage{tikz}
\usetikzlibrary{positioning,arrows.meta,shapes.geometric}
\tikzset{every node/.append style={text depth=0.4ex}}

\patchcmd{\thebibliography}
  {\settowidth}
  {\setlength{\itemsep}{0pt plus 0.1pt}\settowidth}
  {}{}

\title{SERAB: A multi-lingual benchmark for speech emotion recognition}

\name{Neil Scheidwasser-Clow$^{1,2}$*\thanks{*NSC (neil.scheidwasser-clow@epfl.ch) performed this work as an intern at Logitech.}, Mikolaj Kegler$^3$, Pierre Beckmann$^1$, Milos Cernak$^2$}
\address{
  $^1$École Polytechnique Fédérale de Lausanne (EPFL), Lausanne, Switzerland\\
  $^2$Logitech Europe S.A., Lausanne, Switzerland\\
  $^3$Imperial College London, London, United Kingdom}
\begin{document}
\ninept 
\maketitle

\begin{abstract}
Recent developments in speech emotion recognition (SER) often leverage deep neural networks (DNNs). Comparing and benchmarking different DNN models can often be tedious due to the use of different datasets and evaluation protocols. To facilitate the process, here, we present the Speech Emotion Recognition Adaptation Benchmark (SERAB), a framework for evaluating the performance and generalization capacity of different approaches for utterance-level SER. The benchmark is composed of nine datasets for SER in six languages. Since the datasets have different sizes and numbers of emotional classes, the proposed setup is particularly suitable for estimating the generalization capacity of pre-trained DNN-based feature extractors. We used the proposed framework to evaluate a selection of standard hand-crafted feature sets and state-of-the-art DNN representations. The results highlight that using only a subset of the data included in SERAB can result in biased evaluation, while compliance with the proposed protocol can circumvent this issue.
\end{abstract}
\begin{keywords}
emotion recognition, computational paralinguistics, deep neural networks, speech processing, transfer learning
\end{keywords}
\section{Introduction}
\label{sec:intro}

\input{Introduction}

\input{Table1}

\section{Speech Emotion Recognition Adaptation Benchmark (SERAB)}
\input{SERAB}

\section{Baseline approaches}
\label{sec:baseline}
\input{Experiments}

\input{Table3}

\section{Evaluation results \& Discussion}
\input{Results}

\section{Conclusions}
\input{Conclusion}


\newpage
\newpage
\clearpage
\bibliographystyle{IEEEbib}
\bibliography{references}

\end{document}

%% file: Introduction.tex
Speech emotion recognition (SER) is a cornerstone of computational paralinguistics, the analysis of non-verbal elements of speech \cite{schuller2013}. Although a challenging task, being able to automatically and accurately classify emotions from voice could support a wide range of applications, including human-computer interaction~\cite{schuller2013},
healthcare~\cite{van2018}, and public safety~\cite{lefter2011}. At its inception, the development of hand-engineered features has proven effective in tackling various SER-related problems~\cite{schuller2013}. Such features are traditionally based on acoustic~\cite{eyben2010} or linguistic descriptors~\cite{schmitt2017}. More recently, deep neural networks (DNNs) trained in self-supervised fashion were able to produce generalizable representations suitable for a range of audio and speech processing tasks~\cite{shor2020,niizumi2021}. A notable advantage of such data-driven approaches over fixed hand-engineered feature sets is the ability to transfer learned knowledge from a large, unlabeled dataset to downstream tasks with less task-specific data available.

However, the estimated performance and generalization capacity of such self-supervised DNNs can strongly depend on the evaluation protocol. This makes open-source benchmarks, typically composed of fixed dataset(s) and evaluation pipelines, instrumental for informative, fair, and accessible comparisons of different methods. In visual object recognition, ImageNet~\cite{deng2009} has established itself as the \textit{de facto} image dataset and benchmark for deep learning models.
In natural language processing (NLP), GLUE~\cite{wang2018} is a widely used benchmark, with nine different tasks encompassing various characteristics of language understanding (e.g., sentiment analysis, paraphrase, and inference tasks). As one of the largest audio datasets available, AudioSet~\cite{gemmeke2017} is commonly used for self-supervised pre-training, as well as a benchmarking method for audio event classification~\cite{ shor2020, niizumi2021, hershey2017}. A recently proposed HEAR challenge~\cite{hear2021} focuses on evaluating \textit{general-purpose} audio representations and extends the concept underlying AudioSet by including additional tasks. 
In speech representation learning, 
NOSS~\cite{shor2020} was recently proposed as a platform for evaluating speech-specific feature extractors. It includes diverse non-semantic speech processing problems, such as speaker and language identification, as well as two SER tasks (CREMA-D~\cite{cao2014} and SAVEE~\cite{haq2009}).

In contrast to general audio and non-semantic speech representation learning, a standard, readily available multi-task SER benchmark is yet to be introduced. While recently~\cite{fan2021} proposed a SER-specific benchmarking framework, it has two considerable shortcomings. First, it only includes a single dataset, implying the lack of diversity in terms of task difficulty, amount of task-specific data, or data acquisition setup (e.g., recording equipment and conditions). This effectively limits the estimation of generalization capacity, and thus the \textit{real-life} impact of different methods. Second, the dataset is monolingual, with all speech material in English. As a paralinguistic cue, robust embeddings for speech emotion recognition should perform well across different languages.

To that end, we introduce the Speech Emotion Recognition Adaptation Benchmark (SERAB), a collection of nine SER tasks spanning six languages, different dataset sizes and emotion categories. To streamline the comparison of different approaches, we set up a custom evaluation pipeline. We employed the framework to evaluate recent state-of-the-art pre-trained DNNs for speech/audio feature extraction~\cite{shor2020, niizumi2021, hershey2017, plakal2020}, as well as a classic set of hand-crafted features commonly used in computational paralinguistics~\cite{eyben2010}. Lastly, we also propose a novel Transformer-based model, which performs on par with state-of-the-art approaches. Results obtained for a range of baselines demonstrate apparent differences in performance achieved on single datasets, and illustrate the benefits of using the complete SERAB benchmarking framework. 

%% file: Table1.tex
\begin{table*}[t!]
\caption{SERAB tasks and datasets. IEM4: 4-class IEMOCAP \cite{busso2008}. Restricted-access datasets require additional registration on the data provider website. Open-access datasets can be downloaded without registration.}
\vspace{4.5pt}
\centering
\begin{tabular}{cccccccccc}
\toprule
Dataset                        & Code & Access     & Language     & Classes & Utterances & Speakers & Average  & Total     \\
                               &      &            &              &         &         &          & duration (s) & duration (h) \\
\midrule
AESDD \cite{vryzas2018}        & AES  & Open       & Greek        & 5       & 604     & 6        & 4.2  & 0.7          \\
CaFE \cite{gournay2018}        & CAF  & Open       & French       & 7       & 864     & 12       & 4.5  & 1.1          \\
CREMA-D \cite{cao2014}         & CRE  & Open       & English      & 6       & 7,442   & 91       & 2.5  & 5.3          \\
EmoDB \cite{burkhardt2005}     & EMB  & Open       & German       & 7       & 535     & 10       & 2.8  & 0.4          \\
EMOVO \cite{costantini2014}    & EMV  & Open       & Italian      & 7       & 588     & 6        & 3.1  & 0.5           \\
IEM4 \cite{busso2008}          & IEM  & Restricted & English      & 4       & 5,531   & 10       & 3.4  & 7.0           \\
RAVDESS \cite{livingstone2018} & RAV  & Open       & English      & 8       & 1,440   & 24       & 3.7  & 1.5           \\
SAVEE \cite{haq2009}           & SAV  & Restricted & English      & 7       & 480     & 4        & 3.8  & 0.5           \\
ShEMO \cite{nezami2019}        & SHE  & Open       & Persian      & 6       & 3,000   & 87       & 4.0  & 3.3           \\ 
\bottomrule
\end{tabular}
\vspace{-9pt}
\label{tab:serab}
\end{table*}

%% file: SERAB.tex
\subsection{Tasks \& datasets}
A summary of the tasks used in SERAB is presented in Table~\ref{tab:serab}. The benchmark comprises nine speech emotion classification tasks in six languages: four in English (CREMA-D, IEMOCAP, RAVDESS \& SAVEE), and one in French (CaFE), German (EmoDB), Greek (AESDD), Italian (EMOVO), and Persian (ShEMO). In each dataset, speech samples have three attributes: audio data (i.e., the raw waveform, in mono), speaker identifier, and emotion label (e.g., angry, happy, sad). The datasets vary in size (i.e., number of utterances), number of speakers, class distribution, and number of classes. While anger, happiness, and sadness are found across all datasets, disgust, fear, neutral emotion, surprise, calm, and boredom appear in at least one dataset. On the other hand, all datasets have roughly the same average utterance duration (between 2.5 \& 4.5 seconds).

The benchmark was designed to balance dataset popularity, language diversity, and open access. In speech emotion recognition, EmoDB, IEMOCAP and RAVDESS are among the most widely used datasets~\cite{fan2021, nezami2019, xia2015}. In the same vein as~\cite{xia2015}, a 4-class subset of IEMOCAP (IEM4) was used to mitigate the severe class imbalance in the original dataset. For the other tasks, all samples and classes from the original datasets were used  (Table~\ref{tab:serab}). As already present in NOSS~\cite{shor2020}, CREMA-D and SAVEE were included in SERAB. To complete the benchmark, CaFE (French) and EMOVO (Italian) were chosen as Italic-language datasets, whereas AESDD (Greek) and ShEMO (Persian) represented the Hellenic and Indo-Iranian branches of the Indo-European family~\cite{campbell2013}. Overall, the benchmark mainly comprises scripted and acted speech, excepting IEM4~\cite{busso2008}, RAVDESS~\cite{livingstone2018} and ShEMO~\cite{nezami2019} which also feature spontaneous utterances.

Each dataset was split into training, validation, and testing sets to respectively train, optimize and evaluate task-specific speech emotion classifiers.
Excepting CREMA-D, each dataset was split into 60\% training, 20\% validation, and 20\% testing sets. For CREMA-D, we followed a 70/10/20\% (training/validation/testing) split that was applied in NOSS~\cite{shor2020}. 
Each data partition was speaker-independent, i.e., the sets of speakers included in each part were mutually disjoint. Since SERAB datasets vary in size, the fixed data split allows assessing how different methods cope with various amounts of task-specific data.

\subsection{Evaluation pipeline}
\label{sec:eval}

\begin{figure}
    \centering
    \includegraphics[width=0.8\linewidth]{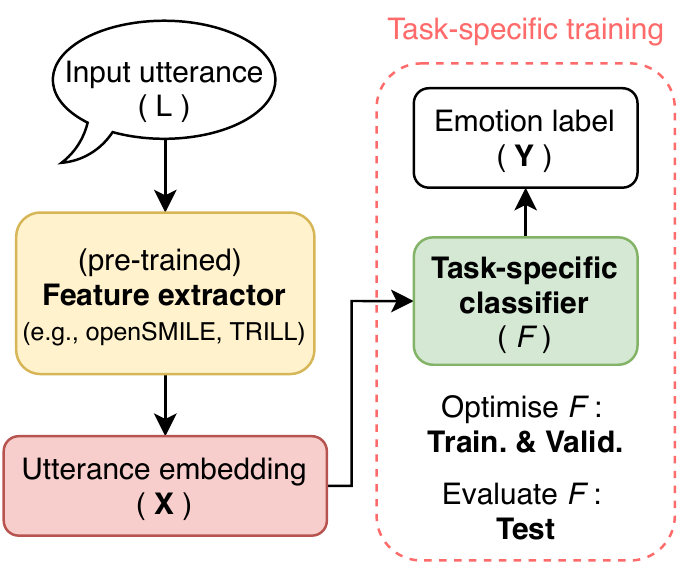}
    \vspace{-9pt}
    \caption{SERAB evaluation pipeline. The (pre-trained/non-trainable) feature extractor is used to obtain utterance-level embeddings (\textbf{X}) from the input. \textbf{X} are used as input to the task-specific classifier $F$ optimized for predicting the emotion \textbf{Y} expressed in the input.}
    \label{fig:evaluation}
    \vspace{-9pt}
\end{figure}

The SERAB evaluation pipeline is used to assess representations of speech-based emotion obtained using different feature extractors (Fig.~\ref{fig:evaluation}). In particular, the workflow includes processing the input utterances through the pre-trained/non-trainable feature extractor and using the resulting embeddings together with a task-specific classifier to predict speech emotion. By using simple classifiers, the classification accuracy reflects the utility of the extracted features for the utterance-level SER tasks. 

Importantly, the utterances included in SERAB vary in duration. The decision about how to integrate information across the utterance is a crucial design choice that may impact feature extractor performance. Thus, we left the decision to the authors of the feature extractor. Most of the current approaches tend to extract information from fixed-length frames and average the outputs across frames. However, other approaches may utilize temporal dependencies in the input utterance. As a result, we require the method to return one set of features for each input utterance with varying duration.

To assess a method's performance, all input utterances are processed through the feature extractor to obtain their embeddings ($\textbf{X}$). The resulting embeddings are then used as features for
basic machine learning classifiers ($F$) trained to predict emotion labels ($\textbf{Y}$). To assure the thorough evaluation, we considered several different classifiers: logistic regression (LR), support vector machine (SVM), linear and quadratic discriminant analysis (LDA/QDA), and random forests (RF). Classifier hyperparameters were optimized through grid-search using the training and validation portions of the data. The best-performing classifier from the grid-search procedure was evaluated on the set-aside test set. All classifier optimization and evaluation procedures were implemented using scikit-learn~\cite{scikit-learn}.

Test-set classification accuracy in each task was used as the performance metric. The resulting accuracies across the nine SERAB tasks were averaged to quantify the method's benchmark performance. In addition to the unweighted mean accuracy (\textbf{UM}) across the SERAB tasks, we also computed the weighted average derived from the test set size (\textbf{WM}), as well as the geometric mean (\textbf{GM}).

%% file: Experiments.tex
\noindent\textbf{openSMILE} - openSMILE~\cite{eyben2010} is a hand-engineered acoustic feature set based on functionals of low-level descriptor contours. Although not directly data-driven, openSMILE is capable of outperforming DNN-based feature extractors, e.g., in problems with little task-specific data~\cite{schuller2021}. 
Here, the most recent implementation of openSMILE\footnote{https://audeering.github.io/opensmile-python/} was used to extract features from each utterance in the SERAB tasks. Subsequently, for each task, the speech emotion classifier was optimized using the training and validation portions of the data and evaluated using the set-aside test set (Section~\ref{sec:eval}). 

\noindent\textbf{VGGish} - VGGish~\cite{hershey2017} is one of the first DNN-based feature extractors for audio, inspired by the VGG-16 convolutional DNN (CNN)~\cite{simonyan2014}. Pre-trained model weights\footnote{https://tfhub.dev/google/vggish/1} were learned through supervised classification of audio events from the Youtube-8M dataset~\cite{abu2016} ($\geq$~350,000 hours of video, 3,000 classes). The model uses fixed-size input windows. To cope with variable-length audio clips, each input utterance was split into non-overlapping $960$ ms-long frames. A log-mel magnitude spectrogram ($N$ = 64 mel frequency bins) was computed from a short-term Fourier transform with 25-ms windows in steps of 10 ms for each frame. The resulting frames were then fed to the pre-trained model for feature extraction. After processing $M$ frames, the obtained $M$ embeddings were averaged to obtain one feature set per utterance. The remaining evaluation followed the protocol outlined in Section~\ref{sec:eval}.

\noindent\textbf{YAMNet} - YAMNet~\cite{plakal2020} is another commonly used DNN-based feature extractor~\cite{shor2020, niizumi2021}. This approach utilizes MobileNetv1~\cite{howard2017mobilenets}, an efficient CNN architecture optimized for mobile devices. Here, we used the weights\footnote{https://tfhub.dev/google/yamnet/1} of the model pre-trained through supervised classification of events from AudioSet~\cite{gemmeke2017} ($\approx$ 5,800 hours of audio, 521 classes). Since the model operates using fixed-size windows, the input utterances were processed analogously to VGGish.

\noindent\textbf{TRILL} - While VGGish and YAMNet were trained on diverse audio sources (speech, music, environmental sounds, etc.), TRILL~\cite{shor2020} was specifically developed as a non-semantic speech feature extractor. The DNN model adopted the architecture of ResNetish~\cite{hershey2017} and was pre-trained in self-supervised fashion using speech samples from AudioSet, which constitutes approximately 50\% of the entire dataset ($\approx$ 2,800 hours of audio). The pre-trained model\footnote{\url{https://tfhub.dev/google/nonsemantic-speech-benchmark/trill/3}} used herein was obtained from triplet loss optimization, which aims at minimizing the embedding-space distance between an anchor and a positive sample (i.e., from the same clip) while maximizing the distance between the same anchor and a negative sample (i.e., from a different clip). In the context of audio, temporally neighboring audio segments will be closer in the representation space and vice versa. Once again, the model operates on fixed-size frames, so the input utterances were processed analogously to VGGish and YAMNet. Following~\cite{shor2020}, we used the embedding from the first 512-depth convolution layer (\textit{layer 19}) which performed best on NOSS.

\noindent\textbf{BYOL-A} - As an alternative to contrastive learning setups such as TRILL, BYOL-A~\cite{niizumi2021} proposes \textit{bootstrapping your own latent} (BYOL) for audio representation learning, inspired by the success of BYOL~\cite{grill2020} for self-supervised image classification. Pre-trained on the entire AudioSet, this approach achieved state-of-the-art results in various audio classification tasks, even outperforming TRILL~\cite{shor2020} in speech processing problems. Instead of assessing the temporal proximity of two different audio segments, BYOL-A relies on comparing two augmented versions of a single sample. More specifically, each version is respectively fed to an \textit{online} network and a \textit{target} network. While both are composed of an encoder and a projection block, the \textit{online} network includes a \textit{prediction} layer which aims at predicting the projected representation of the second augmented view. Thus, BYOL (and BYOL-A) learns a representation by negating the random data augmentations to capture the essential information about the input. Regarding BYOL-A, pre-trained weights for models of different sizes were released by the authors\footnote{https://github.com/nttcslab/byol-a} and used in this work. Since the model accepts inputs of variable length, it returns a single embedding per input utterance. The resulting embeddings are used to train and evaluate the SER classifiers (Section~\ref{sec:eval}).

\noindent\textbf{BYOL-S} - While BYOL-A can achieve state-of-the-art results on a range of audio classification tasks, its \textit{general} audio representation might not be optimal for speech processing and especially paralinguistic problems. Thus, we re-trained BYOL-A using only speech samples of AudioSet, leading to the speech-specific BYOL-S (\textit{S} denoting speech). The model architecture, pre-training routine, and usage remained the same as in the original version.

\noindent\textbf{BYOL-S/CvT} - In this model, we propose an extension of BYOL-S with a Transformer representation. More specifically, we replaced the convolution blocks in BYOL-S with Convolutional Transformer\footnote{https://github.com/lucidrains/vit-pytorch} (CvT)~\cite{wu2021}. CvT notably extends self-attention with depthwise convolution to project the queries, keys, and value embeddings. Between the attention modules, traditional convolution layers are added to decompose the input as in most CNNs. Consequently, CvT combines the qualities of CNNs (e.g., translation invariance) and Transformers (e.g., capturing long-range dependencies and generalization abilities). Here, each CvT stage included only one self-attention layer to allow fair comparisons with BYOL-S, both in terms of model architecture and the number of parameters. We experimented with three different configurations of the model. To explore the impact of model size, the number of filters in CvT stages was manipulated to reduce the number of parameters (Table~\ref{tab:models}), analogously to BYOL-A~\cite{niizumi2021}. In addition, the model was tested with three different embedding dimensions: 256, 512 and 2048. The latter used \textit{mean + max} temporal aggregation in the last layer instead of global average pooling, in the same vein as~\cite{niizumi2021}. Like BYOL-S, the pre-training and application to SERAB tasks was analogous to BYOL-A.

\input{Table2}

%% file: Table2.tex
\begin{table}[t!]
\centering
\caption{Baseline approaches evaluated on SERAB. The embedding size refers to the dimensionatlity of the utterance-level feature set ($\textbf{X}$) used to classify emotional labels ($\textbf{Y}$).}
\vspace{4.5pt}
\setlength\tabcolsep{4.5pt}
\begin{tabular}{lll}
\toprule
Model                                      & Parameters (M)      & Embedding size \\
\midrule
openSMILE~\cite{eyben2010}                  & -    & 6,373        \\
VGGish \cite{hershey2017}                  & 62.0            & 128                      \\
YAMNet \cite{plakal2020}                   & 4.2             & 1,024                    \\
TRILL (layer 19)~\cite{shor2020}            & 9.0             & 12,288 \\
BYOL-A \cite{niizumi2021}                  & 0.6 / 1.6 / 5.3     & 512 / 1,024 / 2,048          \\
\midrule
\textbf{\textit{Proposed:}} \\
BYOL-S                                     & 0.6 / 1.6 / 5.3     & 512 / 1,024 / 2,048          \\
BYOL-S/CvT, \textit{small}           & 1.6             & 256           \\
\footnotesize{\textit{CvT stages: 64/128/256}} \\
BYOL-S/CvT, \textit{large}          & 5.0           & 512 / 2,048       \\
\footnotesize{\textit{CvT stages: 64/256/512}} \\

\bottomrule
\end{tabular}
\vspace{-9pt}
\label{tab:models}
\end{table}

%% file: Table3.tex
\begin{table*}[ht!]
\caption{Test accuracy (\%) on the different downstream tasks in SERAB, referred to by their code from Table~\ref{tab:serab}. \textbf{UM}: unweighted mean, \textbf{WM}: weighted mean (by the number of utterances in the test set), \textbf{GM}: geometric mean. Models are sorted by their \textbf{UM} across all tasks. The best performing approaches for each task and metric are denoted in bold.
}
\vspace{4.5pt}
\label{tab:results}
\centering
\begin{tabular}{llllllllll@{\hspace{26pt}}lll}
\toprule
Model                    & AES  & CAF  & CRE  & EMB  & EMV  & IEM  & RAV  & SAV  & SHE  & \textbf{UM} & \textbf{WM} & \textbf{GM} \\
\midrule
YAMNet                      & 53.6 & 48.1 & 53.9 & 60.7 & 35.7 & 56.1 & 52.3 & 54.2 & 81.7 & 55.1 & 55.8 & 54.0  \\
VGGish                      & 46.4 & 50.0 & 55.5 & 73.8 & 36.2 & 60.1 & 53.0 & 53.3 & 83.6 & 56.9 & 57.7 & 55.4 \\
TRILL, layer 19             & 66.7 & 68.5 & 73.3 & 81.0 & 36.7 & 57.7 & 73.7 & 76.7 & 86.8 & 69.0 & 68.3 & 67.3    \\ 
openSMILE                   & 70.0 & 70.4 & 72.8 & \textbf{90.5} & 37.2 & 62.1 & 71.3 & 72.5 & 84.9 & 70.2 & 69.3 & 68.4 \\       
BYOL-A, 512                 & 71.5 & 73.1 & 70.2 & 84.5 & 39.3 & 62.5 & 74.7 & 76.7 & 90.1 & 72.7 & 69.3 & 69.6   \\ 
BYOL-A, 2048                & 72.0 & 75.5 & 73.7 & 88.1 & 38.3 & 62.8 & \textbf{77.7} & 78.3 & 89.0 & 72.8 & 71.2 & 71.0   \\ 
BYOL-A, 1024                & 75.4 & 74.1 & 71.3 & 88.1 & 44.4 & 62.1 & 76.0 & 80.8 & 89.5 & 73.5 & 70.5 & 72.2   \\ 
\midrule
\textit{\textbf{Proposed:}} \\
BYOL-S/CvT, 256  & 72.9 & 71.8 & 72.9 & 85.7 & 47.4 & 64.8 & 76.0 & 75.8 & 89.0 & 72.9 & 71.5 & 71.9   \\ 
BYOL-S, 512        & 74.9 & \textbf{76.4} & 74.4 & 86.9 & 34.2 & 63.3 & 77.3 & 79.2 & 90.6 & 73.0 & 71.7 & 70.8    \\ 
BYOL-S, 1024       & 75.4 & 72.7 & 75.3 & 84.5 & 39.3 & 63.8 & 74.0 & \textbf{82.5} & 90.9 & 73.2 & 72.1 & 71.5   \\ 
BYOL-S/CvT, 512  & 71.0 & 75.5 & 74.0 & 88.1 & 46.9 & 65.0 & 76.3 & 80.0 & 87.9 & 73.9 & 72.1 & 72.8   \\ 
BYOL-S/CvT, 2048 & 75.8 & 71.3 & \textbf{76.9} & 84.5 & \textbf{48.5} & \textbf{65.1} & 76.3 & 76.7 & \textbf{93.0} & 74.2 & \textbf{73.6} & 73.2   \\ 
BYOL-S, 2048       & \textbf{77.3} & 74.5 & \textbf{76.9} & 88.1 & 44.4 & 64.8 & 76.7 & 81.7 & 91.1 & \textbf{75.1} & \textbf{73.6} & \textbf{73.7} \\ 
\bottomrule
\end{tabular}
\vspace{-9pt}
\end{table*}

%% file: Results.tex
Table~\ref{tab:models} presents configurations of different baseline approaches outlined in Section~\ref{sec:baseline}. For BYOL-like models (BYOL-A, -S, -S/CvT), we explored different model sizes and sizes of the output embedding fed to the task-specific classifiers predicting emotion labels. Benchmark performance for all baseline methods is presented in Table~\ref{tab:results}. 

The large BYOL-S, with a 2048 embedding size, emerged as the best model across all considered performance metrics and yields the best individual accuracy on two out of the nine SERAB tasks. Importantly, model ranks were generally similar across all benchmark-wide metrics. Thus, we chose to sort model performance by \textbf{UM} in accordance with previous benchmarks for computer vision systems~\cite{zhai2019}. Although reaching slightly lower scores across the entire benchmark, the largest BYOL-S/CvT remained competitive by providing the best results in four out of nine tasks. Moreover, the increase in the embedding size and the overall model size tend to consistently improve the proposed approaches' performance.

More generally, all BYOL-inspired models, even with small sizes, achieved significantly higher scores (up to a 5\% absolute difference in \textbf{UM}) than TRILL, VGGish, YAMNet, and openSMILE. This considerable difference in performance most likely originates from the vastly different pre-training strategies. Another reason for the supremacy of BYOL-derived models might be the fact that they are designed to process variable-length inputs rather than fixed-length frames as openSMILE, VGGish, YAMNet, and TRILL do. This, in turn, suggests that aggregation of the temporal context could improve utterance-level SER performance.

Interestingly, BYOL-S models performed consistently better than original BYOL-A approaches. This indicates that specializing the pre-training task by focusing only on speech excerpts resulted in more suitable embeddings for SER. In such a speech-specific pre-training, the model presumably developed better capacity for representing speech, including language-independent paralinguistic cues such as speech-based emotion.

On the other hand, enriching BYOL-S with self-attention mechanisms via CvT did not yield a notable performance increase we anticipated, but the model was lighter with 0.3M fewer parameters than BYOL-S. The slight difference in terms of performance might be due to the minimal inductive biases implied in Transformer-like models, in contrast to CNNs~\cite{cazenavette2021}. While advantageous when training large models on large datasets~\cite{dosovitskiy2020}, such biases become critical in smaller network and smaller dataset setups~\cite{cazenavette2021}. Thus, increasing the pre-training dataset's size could help develop the generalization ability of Transformers and thus improve the overall performance of BYOL-S/CvT.

While SERAB allows comparing different models across a diverse range of tasks, more importantly, it provides a streamlined benchmarking platform for the comparison of different approaches. In particular, some tasks exhibit significant variations between models (e.g., EMOVO, SAVEE, EmoDB) such that the overall poorer-performing approaches may appear better than they really are. Some of these differences might be dataset-specific, introducing even larger bias that is not trivial to overcome. The inclusion of multiple tasks across different languages provides robust performance estimates, as shown by our evaluation of the baseline approaches.

%% file: Conclusion.tex
We introduce SERAB, a multi-lingual benchmark for speech emotion recognition. With the rapid emergence of DNN-based representations of speech and speech-based emotion, the benchmark provides a universal platform for comparing different methods. Due to the inclusion of diverse tasks spanning across different languages, dataset sizes, and emotional categories, SERAB produces robust estimates of performance and generalization capacity. We used SERAB to evaluate a range of recent baselines. 
Among the tested frameworks, BYOL-based approaches yielded superior performance across all considered metrics. Interestingly, pre-training BYOL-A models on only speech samples of AudioSet (BYOL-S) led to an almost 3\% accuracy improvement compared to the original method. Presented evaluation results can be used as baselines for developing novel approaches, such as CvT-based methods explored here. Future work should focus on incorporating more datasets in even more languages into SERAB, as well as extending the task range to include regression problems such as valence or arousal estimation. To facilitate the usage of SERAB, the framework, including setup instructions, evaluation pipelines \& examples, is freely available online\footnote{\url{https://github.com/Neclow/serab/}}.

